\documentstyle[aps,preprint,tighten]{revtex}
\begin{document}

\draft
\input{epsf}
\preprint{TUTP-97-10, gr-qc/9710055}

\title{Scalar Field Quantum Inequalities
        in Static Spacetimes}

\author{Michael J. Pfenning\footnote{email: mitchel@cosmos2.phy.tufts.edu}
and L. H. Ford\footnote{email: ford@cosmos2.phy.tufts.edu}}

\address{Institute of Cosmology, Department of Physics and
Astronomy, Tufts University, Medford, Massachusetts 02155, USA}

\date{October 9, 1997}

\maketitle

\begin{abstract}
We discuss quantum inequalities for minimally coupled scalar fields in
static spacetimes.  These are inequalities which place limits on the
magnitude and duration of negative energy densities.  We derive a
general expression for the quantum inequality for a static observer
in terms of a Euclidean two-point function.  In a short sampling time
limit, the quantum inequality can be written as the
flat space form plus subdominant correction terms dependent
upon the geometric properties of the spacetime.  This supports the
use of flat space quantum inequalities to constrain negative energy
effects in curved spacetime.  Using the exact Euclidean two-point
function method, we develop the quantum inequalities for perfectly
reflecting planar mirrors in flat spacetime.  We then look at the
quantum inequalities in static de~Sitter spacetime,  Rindler spacetime
and two- and four-dimensional black holes.  In the case of a
four-dimensional Schwarzschild black hole, explicit forms of the
inequality are found for static observers near the horizon and at large
distances.  It is show that there is a quantum averaged weak energy
condition (QAWEC), which states that the energy density averaged over
the entire worldline of a static observer is bounded below by the vacuum
energy of the spacetime. In particular,
for an observer at a fixed radial distance away from a black hole, the QAWEC 
says that the averaged energy density can never be less than the
Boulware vacuum energy density.
\end{abstract}

\pacs{04.62.+v, 03.70.+k, 11.10.-z, 04.60.-m}


\section{Introduction}

In a recent paper \cite{Pfen97a}, we derived a general form of the
quantum inequality (QI) for quantized scalar fields in static curved
spacetimes.  The quantum inequalities are uncertainty type relations
which constrain the magnitude and duration of negative energy that
may be present in a spacetime.  This was an extension of the previous
work carried out by Ford and Roman \cite{Ford91,F&Ro92,F&Ro95,F&Ro97}
which dealt with the quantum
inequalities for scalar fields in two- and four-dimensional
Minkowski spacetime.   In curved spacetimes, it was found that
the quantum inequality could be written in terms of a sum of
mode functions for the scalar field.  With the general form of
the quantum inequality in hand, we then proceeded to look at the static
cases of the three-dimensional closed universe and the four-dimensional
Robertson-Walker spacetimes.  Exact functional forms for the quantum
inequalities were developed in these spacetimes.  It was found that
the curved space quantum inequalities could be written as the flat
space quantum inequalities multiplied by a ``scale'' function which
detailed the behavior of the inequalities at various ratios of the
sampling time to the radius of curvature of the spacetime.
In the long sampling time limit, the quantum inequality
was substantially modified by the scale function.  However, in
the short sampling time limit, the scale functions tend to 1,
yielding the flat space quantum inequality.  This behavior had first
been predicted to exist by Ford and Roman in a paper dealing with
negative energy around wormholes \cite{F&Ro96}.  It was argued that
by making the sampling time of the quantum inequality much shorter
than a minimum characteristic curvature scale, then the spacetime
could be considered locally flat and the Minkowski space quantum inequality
should hold.  This method has since been applied to the Alcubierre
``Warp Drive'' metric \cite{Alcu94,Pfen97b} to show that the negative
energy that makes up the walls of the warp bubble has to be constrained
to exceptionally thin walls, usually on the order of hundreds, or perhaps
thousands of Planck lengths at most.  Similar results were found for the
Krasnikov metric \cite{Kras95,E&Ro97}, where the negative energy that
is needed must also be confined to exceptionally thin walls.   

In our earlier work \cite{Pfen97a},  the quantum inequality was derived
for stationary observers in static spactimes where the magnitude of the
$g_{tt}$ component of the metric was 1.  In this paper we will extend the
derivation of the quantum inequalities to the entire class of static
spacetime metrics of the form
\begin{equation}
{ds}^2 = -|g_{tt}({\bf x})|{dt}^2 + g_{ij}({\bf x})dx^i dx^j \;.
\label{eq:metric}
\end{equation}
In Sec.~\ref{sec:QI} we will derive the general form of the quantum
inequality for static observers in such spacetimes, and show that it
may be written in terms of a Euclidean Green's (two-point) function.
We will show in Sec.~\ref{sec:QAWEC} that in the infinite sampling
time limit the quantum inequality reduces to the ``quantum averaged weak
energy condition'' (QAWEC) which can be written in the form
\begin{equation}
\int_{-\infty}^\infty \left( \langle \psi | T_{00}/|g_{tt}|\, | \psi \rangle
- \rho_{vac} \right) d\tau \geq 0 \, .
\end{equation}
The quantum averaged weak energy condition says that along the entire
world-line of a static observer, the sampled energy density can
never be more negative than the vacuum energy, $\rho_{vac}$.  Here
the vacuum energy is obtained using the timelike killing vector to
define positive frequency.

In Sec.~\ref{sec:expans} we will perform a short time expansion of
the two-point function. It is found that the leading term of the
expansion of the curved space quantum inequality is indeed of the
flat space form. In addition, the first two corrections to the leading
order term will be explored.  We will show that they depend only on the
geometric properties of the spacetime such as the metric,
scalar curvature, etc. 

In Sec.~\ref{sec:planar} we will look at the exact form of the
quantum inequality developed for a half infinite flat spacetime.  We
will see that the presence of a perfectly reflecting, infinite planar
mirror modifies the flat space quantum inequality.  In addition we
will look at the case of the quantum inequality between two parallel
mirrors.  Both of these cases will be developed by first determining
the Feynman Green's function by the method of images, and then using
the formalism developed in Sec.~\ref{sec:QI} to find the respective
quantum inequalities.

Finally, we will look at the quantum inequalities in spacetimes
in which there exist horizons.  We will begin with the two-dimensional
Rindler coordinates and then move on to the static coordinate representation
of de~Sitter spacetime.  Finally in Sec.~\ref{sec:black_holes} we
will look at the case of two- and four-dimensional black holes.
In two-dimensions, we will find the exact form of the quantum inequality
for static observers sitting at fixed radii outside of the black hole.
In the case of the four dimensional black hole, because there is no
known analytic solution for the mode functions of the scalar field, we
find the quantum inequality in the limits $r\rightarrow 2M$ and 
$r\rightarrow\infty$. In the limit of long
sampling time, the QAWEC is recovered for these spacetimes.  


\section{The Scalar Field Quantum Inequality}\label{sec:QI}

Because this derivation closely resembles that developed earlier
\cite{Pfen97a}, we will only highlight the necessary steps to replicate
the proof for the metric in Eq.~(\ref{eq:metric}). On such a fixed
background, the wave equation 
\begin{equation}
\Box\phi - m^2\phi  = 
{1\over \sqrt{|g|}}\partial_\alpha \left(\sqrt{|g|} g^{\alpha\beta}
\partial_\beta \phi\right)- m^2\phi = 0
\end{equation}
becomes
\begin{equation}
 -{1\over |g_{tt}|}\partial_t^2\phi +
\nabla^j \nabla_j \phi - m^2\phi = 0,
\end{equation}
where $g={\rm det}(g_{\mu\nu})$, $\nabla_i$ is the covariant derivative
in the spacelike hypersurfaces orthogonal to the Killing vector, 
and $m$ is the mass of the field.  Units where $\hbar = c = G = 1$ 
are used throughout this paper.  The positive frequency mode
function solutions can be written as
\begin{equation}
f_\lambda({\bf x},t) = \,U_\lambda ({\bf x}) \, e^{-i\omega t},
\end{equation}
where $U_\lambda({\bf x})$ is the solution to the Helmholtz equation
\begin{equation}
\nabla^j\nabla_j U_\lambda +(\omega^2_\lambda/|g_{tt}| - m^2)U_\lambda =0\, .
\label{eq:Helmholtz}
\end{equation}
The label $\lambda$ represents the set of quantum numbers necessary to
specify the mode. Additionally, the mode functions are defined to have unit 
Klein-Gordon norm. A general solution of the scalar field $\phi$ can then be 
expanded in terms of creation and annihilation operators as
\begin{equation}
\phi(x) = \sum_{\lambda} {\bigl(a_{\lambda}f_{\lambda} + a^\dagger_{\lambda}
f^\ast_{\lambda}\bigr)},
\end{equation}
where quantization is carried out over a finite box or universe.  If the
spacetime has infinite spatial extent, then we replace the summation by
an integral over all of the possible modes.

In the development of the quantum inequality, we will concern ourselves
only with static observers, whose four-velocity, $u^\mu = ( |g_{tt}|^{-1/2}
 , {\bf 0 })$, is parallel to the direction of the timelike
Killing vector. These are geodesic observers in the case that $g_{tt}$ is
a constant, but otherwise are non-geodesic.  The energy density (for
minimal coupling) that such an observer measures is given by
\begin{equation}
\rho = T_{\alpha\beta} u^\alpha u^\beta = {1\over |g_{tt}|}T_{00}
= {1\over 2}\left[ {1\over |g_{tt}|}(\partial_t\phi)^2 +\nabla^j\phi
\nabla_j\phi + m^2\phi^2\right]\,. \label{eq:rho}
\end{equation}
Upon substitution of the above mode function expansion into 
Eq.~(\ref{eq:rho}), one finds that there exists a vacuum energy
term which is divergent upon summation.  A regularization and
renormalization scheme is needed to define the physical energy
density.  This may be side-stepped
by concentrating attention upon the difference between the energy
density in an arbitrary state and that in the vacuum state, as was
done in Ref.~\cite{Pfen97a,F&Ro95}. We will therefore concern ourselves
primarily with the normal ordered quantity
\begin{equation}
:\rho:\; = \rho - \langle 0 |\rho | 0\rangle,
\end{equation}
where $| 0\rangle$ represents the Fock vacuum state defined by
the global timelike Killing vector.  In cases where the renormalized 
value of $\langle 0 | \rho | 0\rangle$ is known, we can convert the
difference inequality into an inequality on the renormalized energy
density in an arbitrary state. 

The energy density as defined above is valid along the
entire worldline of the observer.  However, let us sample the
energy density  only along some finite interval of the
geodesic.  This may be accomplished by means of a weighting function
which has a characteristic time $t_0$,  such as the Lorentzian function,
\begin{equation}
h(t) = {t_0 \over\pi}\,{1\over {t^2 +t_0^2}}. 
\end{equation}
The integral over all time of $h(t)$ is equal to one
and the width of the Lorentzian is characterized by $t_0$. Using such
a weighting function, one finds that the averaged energy difference is
given by
\begin{eqnarray} 
\Delta\hat \rho &\equiv& {t_0\over\pi}\int_{-\infty}^\infty
{{\langle :T_{00}/|g_{tt}|:\rangle dt} \over{t^2 + t_0^2}}\nonumber\\
& = & {\rm Re}\sum_{\lambda\lambda'}\left\{{{\omega\omega'}\over |g_{tt}|}
\left[ U_\lambda^* U_{\lambda'} e^{-|\omega-\omega'|t_0}\langle 
a_{\lambda}^\dagger a_{\lambda'}\rangle - U_\lambda U_{\lambda'} 
e^{-(\omega+\omega')t_0}\langle a_{\lambda} a_{\lambda'}\rangle\right]
\right.\nonumber\\
&& \qquad + \left[ \nabla^j U_\lambda^* \nabla_j 
U_{\lambda'} e^{-|\omega-\omega'|t_0}\langle a_{\lambda}^\dagger
a_{\lambda'}\rangle+\nabla^j U_\lambda \nabla_j U_{\lambda'}
e^{-(\omega+\omega')t_0}\langle a_{\lambda} a_{\lambda'}\rangle\right]
\nonumber\\
& &\qquad + m^2 \left.\left[ U_\lambda^* U_{\lambda'} 
e^{-|\omega-\omega'|t_0}\langle a_{\lambda}^\dagger a_{\lambda'}\rangle 
+  U_\lambda U_{\lambda'} e^{-(\omega+\omega')t_0}\langle a_{\lambda} 
a_{\lambda'}\rangle\right] \right\}. \label{eq:DeltaRho}
\end{eqnarray}
From this point onward, the derivation continues along the lines of
that in Ref.~\cite{Pfen97a}.  After some algebra, and application
of the inequalities derived in previous papers \cite{Pfen97a,F&Ro97},
one finds
\begin{equation}
\Delta\hat \rho \geq - \sum_\lambda \left({\omega_\lambda^2\over |g_{tt}|}
 + {1\over 4}\nabla^j\nabla_j\right) |U_\lambda({\bf x})|^2 {\rm e}^{-2
\omega_\lambda t_0}\, ,                           \label{eq:qi2}
\end{equation}
which can be rewritten as
\begin{equation}
\Delta\hat \rho \geq - {1\over 4}
\left({\partial_{t_0}^2\over |g_{tt}|} + \nabla^j\nabla_j\right)
\sum_\lambda |U_\lambda({\bf x})|^2 {\rm e}^{-2\omega_\lambda t_0}\, .
\label{eq:general_qi}
\end{equation}
There is a more compact notation in which Eq.~(\ref{eq:general_qi})
may be expressed.  If we take the original metric, Eq.~(\ref{eq:metric}),
and Euclideanize the time by allowing $t \rightarrow it_0$ then the
Euclidean box operator is defined by
\begin{equation}
\Box_E \equiv {\partial_{t_0}^2\over |g_{tt}|} + \nabla^j\nabla_j\, .
\end{equation}
In addition, the sum of the mode functions is equal to
the Euclidean two-point function
\begin{equation}
G_E({\bf x},-t_0;{\bf x},+t_0) = \sum_\lambda |U_\lambda({\bf x})|^2 
{\rm e}^{-2\omega_\lambda t_0} 
\end{equation}
where the spatial separation is allowed to go to zero but the time
separation is $2t_0$.  The Euclidean two-point function is the
counterpart of the Feynman Green's function for the Lorentzian 
metric.  The two are related by
\begin{equation}
G_E({\bf x},t;{\bf x}',t') = i G_F({\bf x},it;{\bf x}',it').
\label{eq:Euclideanize}
\end{equation}
This allows us to write the quantum inequality in any static curved
spacetime as
\begin{equation}
\Delta\hat \rho \geq - {1\over 4}\Box_E \, G_E({\bf x},-t_0;{\bf x},+t_0).
\label{eq:QI}
\end{equation}
We see that once we are given a metric which admits a
timelike Killing vector, we can calculate the limitations on
the negative energy densities by either of two methods.  If
we know the solutions to the wave equation, then we may construct
the inequality from the summation of the mode functions. More
elegantly, if the Feynman two-point function is known in the
spacetime, then we may immediately calculate the inequality by
first Euclideanizing and then taking the appropriate derivatives. 


\section{The Quantum Averaged Weak Energy Condition}\label{sec:QAWEC}
Let us return to the form of the the quantum
inequality given by Eq.~(\ref{eq:qi2}),
\begin{equation}
\Delta\hat \rho \geq - \sum_\lambda \left({\omega_\lambda^2\over |g_{tt}|}
 + {1\over 4}\nabla^j\nabla_j\right) |U_\lambda({\bf x})|^2 {\rm e}^{-2
\omega_\lambda t_0}\, .                           
\end{equation}
Since we are working in static spacetimes, the vacuum energy 
does not evolve with time, so we can rewrite this equation simply
by adding the renormalized vacuum energy density $\rho_{vacuum}$
to both sides. We then have
\begin{equation}
\hat\rho_{Ren.} \geq - \sum_\lambda \left({\omega_\lambda^2\over |g_{tt}|}
 + {1\over 4}\nabla^j\nabla_j\right) |U_\lambda({\bf x})|^2 {\rm e}^{-2
\omega_\lambda t_0} + \rho_{vacuum}({\bf x})\, ,                           
\end{equation}
where $\hat\rho_{Ren.}$ is the sampled, renormalized energy density
in any quantum state.  Let us now take the limit of the sampling time
$t_0 \rightarrow \infty$. We find (under the assumption that there exist
no modes which have $\omega_\lambda = 0$) that
\begin{equation}
\lim_{t_0\rightarrow\infty} {t_0\over\pi} \int_{-\infty}^\infty 
{\langle T_{00}/|g_{tt}| \rangle_{Ren.} \over  t^2 + t_0^2} dt \geq
 \rho_{vacuum}({\bf x}), 
\end{equation}
This leads directly to the ``quantum averaged weak energy condition''
for static observers,
\begin{equation}
\int_{-\infty}^{+\infty} \left( \langle \psi | T_{00}/|g_{tt}| 
|\psi\rangle_{Ren.} - \rho_{vacuum} \right) dt \geq 0 .
\label{eq:QAWEC}
\end{equation} 
This is a departure from the classical averaged weak energy condition,
\begin{equation}
\int_{-\infty}^{+\infty} \langle \psi | T_{00}/|g_{tt}|\,
 |\psi\rangle_{Ren.} dt \geq 0 .
\end{equation}
We see that the derivation of the QAWEC leads to the measured energy
density along the observers geodesic being bounded below by the vacuum
energy.  Recently, there has been much discussion about how
badly the vacuum energy violates the classical energy conditions.
For example Visser looked at the specific case of the violation
of classical energy conditions for the Boulware, Hartle-Hawking, and
Unruh vacuum states \cite{Viss96a,Viss96b,Viss96c,Viss97a} around a 
black hole.  However the vacuum energy is not a classical phenomenon,
so it necessarily need not obey classical energy constraints.  From the
QAWEC we see that the sampled energy density is bounded below
by the vacuum energy in the long sampling time limit.  


\section{Expansion of the QI for Short Sampling Times}\label{sec:expans}

We now consider the expansion of the two-point function
for small times.  We assume that the two-point function has the
Hadamard form \cite{B&OT86}
\begin{equation}
G(x, x') = {i\over 8\pi^2} \left[ {\Delta^{1/2} \over \sigma + i\epsilon}
+ V \ln(\sigma + i\epsilon) +W\right]\, ,
\end{equation}
where $2\sigma(x, x')$ is the square of the geodesic distance between
the spacetime points $x$ and $x'$,
\begin{equation}
\Delta\equiv - g^{-1/2}(x)\, {\rm det}(\sigma_{;ab'})\: g^{-1/2}(x'),
\end{equation}
is the Van Vleck-Morette determinant, and $V(x, x')$ and $W(x, x')$ 
are regular biscalar functions. In general, these functions can be
Taylor series expanded in powers of $\sigma$,
\begin{equation}
V(x,x') = \sum_{n=0}^\infty V_n(x,x') \sigma^n \, ,
\end{equation}
where $V_n$ (and $W_n$) is also a regular biscalar function with
\begin{eqnarray}
V_0 &=& v_0 - {1\over 2} v_{0;a}\,\sigma^a + {1\over 2}v_{0ab}\,\sigma^a 
\sigma^b + {1\over 6} (-{3\over 2}v_{0ab;c}+ {1\over 4}v_{0;(abc)})\sigma^a 
\sigma^b \sigma^c +\cdots ,\\[8pt]
V_1 &=& v_1 - {1\over 2} v_{1;a}\,\sigma^a + \cdots,
\end{eqnarray}
where $\sigma^\nu = \sigma^{;\nu}$.  The coefficients, $v_0$, $v_{0ab}$,
\ldots are strictly geometrical objects given by
\begin{eqnarray}
v_0 &=& {1\over 2} \left[ (\xi - {1\over 6}) R + m^2 \right] ,\\[8pt]
v_{0ab} &=& -{1\over 180} R_{pqra} {R^{pqr}}_{b}  -{1\over 180} R_{apbq}R^{pq}
+{1\over 90} R_{ap}{R_b}^p - {1\over 120}\Box R_{ab} \nonumber\\
&&+ ({1\over 6}\xi -
{1\over 40}) R_{;ab} + {1\over 12}(\xi - {1\over 6}) R R_{ab} +
{1\over 12} m^2 R_{ab}\, ,\\[8pt]
v_1 &=& {1\over 720} R_{pqrs}R^{pqrs} - {1\over 720} R_{pq}R^{pq}
-{1\over 24}(\xi - {1\over 5})\Box R + {1\over 8}(\xi -{1\over 6})^2 R^2
\nonumber\\ &&+ {1\over 4}m^2 (\xi-{1\over 6})R + {1\over 8} m^4.
\end{eqnarray}
We can then express the Green's function as
\begin{eqnarray}
G(x,x') &=& {i\over 8\pi^2} \left\{ {1+{1\over 12}R_{ab}\sigma^a \sigma^b\
- \cdots \over \sigma + i\epsilon} + \left[ (v_0 - {1\over 2}v_{0;a}
\sigma^a + {1\over 2}v_{0ab} \sigma^a \sigma^b\ + \cdots ) \right. \right. 
\nonumber\\ &&\qquad \left. \left. +( v_1 -
{1\over 2} v_{1;a}\sigma^a + \cdots)\sigma + \cdots \right] \ln 
( \sigma + i\epsilon ) + W \right\}
\end{eqnarray}
where we have also used the Taylor series expansion of the 
Van Vleck-Morette determinant \cite{B&OT86},
\begin{equation}
\Delta^{1/2} = 1 + {1\over 12}R_{ab}\sigma^a \sigma^b
- {1\over 24}R_{ab;c}\sigma^a \sigma^b \sigma^c + \cdots \, .
\end{equation}
We neglect $W$, the state dependent part of the Green's function,
because it is regular as $\sigma \rightarrow 0$.  The dominant
contributions to the quantum inequality come from the divergent
portions of the Green's function in the $\sigma \rightarrow 0$
limit.

\begin{figure}
\begin{center}
\leavevmode\epsffile{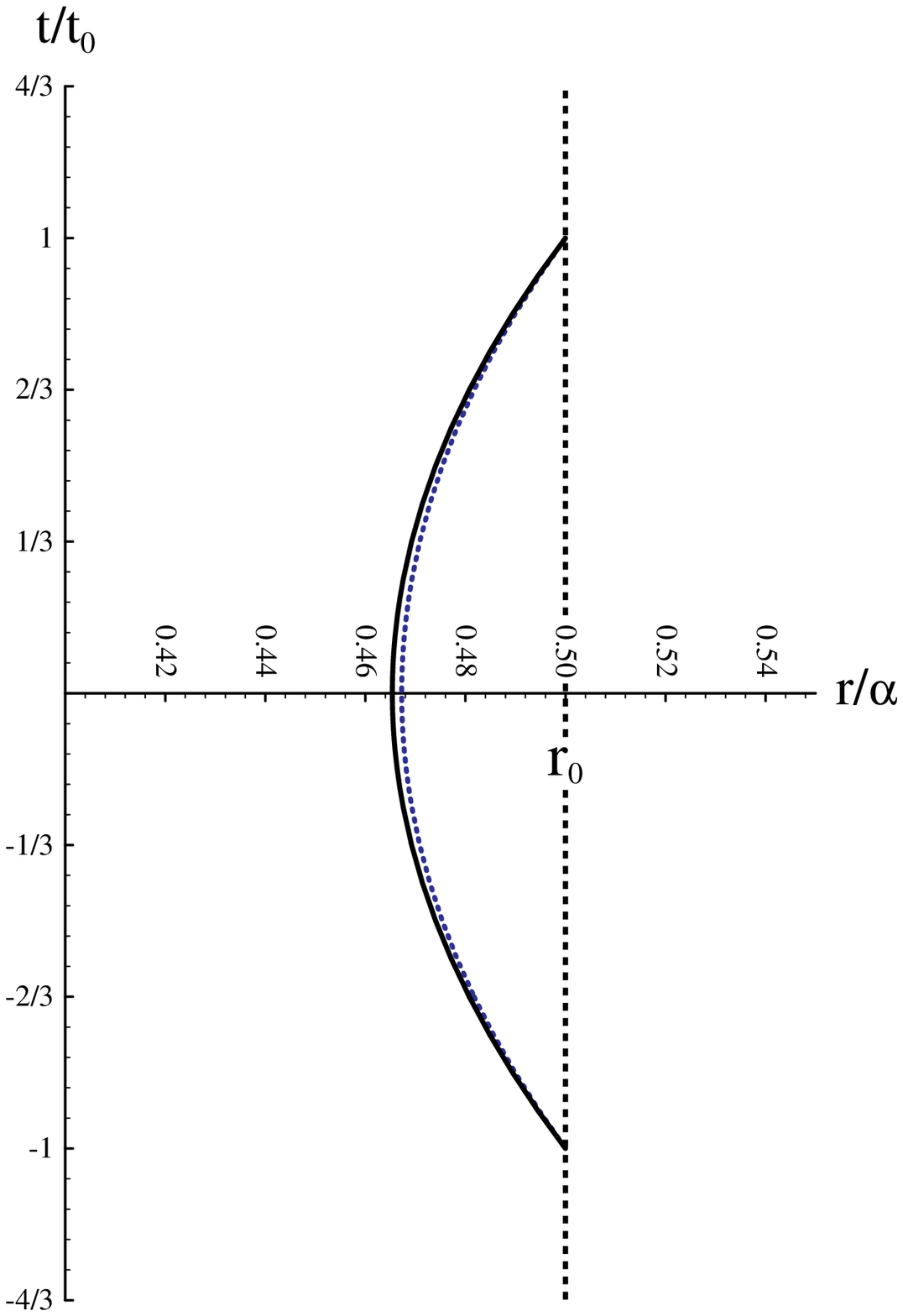}
\end{center}
\caption[Exact and approximate geodesics in de~Sitter spacetime]
{ An exaggerated plot of the exact geodesic path (dotted line) and
the parabolic approximation (solid line).  $\alpha$ is the coordinate distance
from $r=0$ to the horizon in the static de~Sitter spacetime. }
\label{fig:geodesicpath}
\end{figure}

Let us find the geodesic distance between two spacetime
points, along a curve starting at $({\bf x}_0,-t_0)$ and ending at
$({\bf x}_0,+t_0)$.  For spacetimes in which $|g_{tt}| = 1$,
the geodesic path between these two spacetime points is a straight line.
Therefore, the geodesic distance is simply $2t_0$.  However, in a more
generic static spacetime where $g_{tt}({\bf x})$ is not constant,
the geodesic path between the above two spacetime points is a curve,
with the observer's spatial position changing throughout time.  Thus,
we must now solve the equations of motion for the observer.  In terms
of an affine parameter $\lambda$, the geodesic equations are found to be
\begin{eqnarray}
{dt\over d\lambda} - {a_t \over |g_{tt}({\bf x}(\lambda) )|} & = & 0 \, ,\\
{d^2x^i \over d\lambda^2} + {\Gamma^i}_{\mu\nu} {dx^\mu \over d\lambda}
{dx^\nu \over d\lambda} & = & 0 \; ,
\end{eqnarray}
where $a_t$ is an unspecified constant of integration.  The Christoffel
coefficients are
\begin{eqnarray}
{\Gamma^i}_{tt} & = & {1\over 2} g^{ij}\, |g_{tt}|_{,j} \, ,\cr
{\Gamma^i}_{tj} & = & 0 \, ,\cr
{\Gamma^i}_{jk} & = & {1\over 2} g^{im}\, \left( g_{mj,k} + g_{mk,j} -
g_{jk,m} \right)\, .
\end{eqnarray}
It is possible to eliminate $\lambda$ from the position equations,
and write
\begin{equation}
{d^2 x^i \over dt^2} + {1\over 2}  |{g_{tt}}|^{,i} + 
{\Gamma^i}_{jk} {dx^j \over dt} {dx^k \over dt} +  {|g_{tt}|_{,k} \over
|g_{tt}| }{dx^i \over dt} {dx^k \over dt} = 0.
\end{equation}
Now if we make the assumption that the velocity of the observer moving
along this geodesic is small, then to lowest order the second term can
be considered nearly constant, and all the velocity dependent terms are
neglected.  It is then possible to integrate the equation exactly,
subject to the above endpoint conditions, to find 
\begin{equation}
x^i(t) \approx  -{1\over 4} {|g_{tt}|}^{,i}_{{\bf x} = {\bf x_0}} 
(t^2 -t_0^2) + x_0^i \, .
\label{eq:geodesicpath}
\end{equation}
We see that the geodesics are approximated by parabolae, as would be 
expected in the Newtonian limit.  A comparison of the exact solution to
the geodesic equations and the approximation is shown in
Fig.~\ref{fig:geodesicpath} for the specific case of de~Sitter spacetime.
We see that the approximate path very nearly fits the exact path in the
range of $-t_0$ to $+t_0$.

The geodesic distance between two spacetime points,
where the starting and ending spatial positions are the same,
is given by
\begin{equation}
\Delta s = \int_{-t_0}^{+t_0} \sqrt{ -|g_{tt}(t)| + g_{ij}(t) {dx^i\over dt}
{dx^j\over dt}} dt
\end{equation}
In order to carry out the integration, let us define 
\begin{equation}
f(t) \equiv \sqrt{ -|g_{tt}(t)| + g_{ij}(t) {dx^i\over dt}{dx^j\over dt}}.
\end{equation}
We can expand $f(t)$ in powers of $t$ centered around $t = 0$, 
and then carry out the integration to find the geodesic distance.
The parameter $\sigma$ can now be written as
\begin{eqnarray}
\sigma({\bf x}_0,t_0) & = &{1\over 2}\, {\Delta s}^2\\
& = & 2 f^2(0)\, t_0^2 + {2\over 3}
f(0)f^{''}(0) \, t_0^4 + {1\over 6}\left[ {1\over 5} f(0) f^{(IV)}(0) + 
{1\over 3} f^{''}(0)^2\right] \, t_0^6 + \cdots.
\end{eqnarray}
However, we do not necessarily know the values of the metric at the 
time $t = 0$, but we do at the initial or final positions, so we must
now expand the functions $f(0)$ around the time $-t_0$.
Upon using Eq.(\ref{eq:geodesicpath}), one then finds that
\begin{equation}
\sigma({\bf x_0},t_0) \approx -2 |g_{tt}({\bf x_0})|\, t_0^2 - {1\over 6}\,
{g_{tt}}^{,i}({\bf x_0}) \, g_{tt,i}({\bf x_0})\, t_0^4+\cdots
\end{equation}
and
\begin{equation}
\sigma^t({\bf x_0},t_0) \approx 2 t_0 + {1\over 3}
{{g_{tt}}^{,i}({\bf x_0}) \, g_{tt,i}({\bf x_0}) \over
|g_{tt}({\bf x_0})|} \, t_0^3 + \cdots.
\end{equation}
In any further calculations, we will drop the ${\bf x_0}$ notation, with
the understanding that all of the further metric elements are evaluated at
the starting point of the geodesic.  
Using Eq.~(\ref{eq:Euclideanize}) we can then write the Euclidean
Green's function needed to derive the quantum inequality, in increasing
powers of $t_0$ as
\begin{eqnarray}
G_E(x,t_0) &=& {1\over 8 \pi^2} \left[ { {1 - O(t_0^2) + \cdots}\over 
{2 |g_{tt}|\, t_0^2 - {1\over 6} {g_{tt}}^{,i} g_{tt,i}\, t_0^4}}
+ v_0 \ln ( 2 |g_{tt}|\, t_0^2 - {1\over 6} {g_{tt}}^{,i} g_{tt,i}\,
t_0^4)\right.\nonumber\\ 
&&+ \left.\left( v_{0,k}\, |{g_{tt}}|^{,k} +  2 v_1 |g_{tt}|
- 2 v_{000} \right) t_0^2 \ln ( 2 |g_{tt}|\, t_0^2 - {1\over 6}
{g_{tt}}^{,i} g_{tt,i}\, t_0^4)+\cdots \right]
\end{eqnarray} 
Note, none of the geometric terms, such as $v_0$, change during 
Euclideanization because they are time independent.  The
quantum inequality, (\ref{eq:QI}), can be written as
\begin{equation}
\Delta\hat \rho \geq - {1\over 4} \left( {1\over |g_{tt}|} 
\partial_{t_0}^2 + \nabla^i \nabla_i \right) \,
 G_E({\bf x},-t_0;{\bf x},+t_0). 
\end{equation}
If we insert the Taylor series expansion for the Euclidean Green's
function into the above expression and collect terms in powers of
the proper sampling time $\tau_0$, related to $t_0$
by $\tau_0 = |g_{tt}|^{1/2} t_0$, we can write the above expression as
\begin{eqnarray}
\Delta\hat\rho &\geq& -{3\over 32 \pi^2 \tau_0^4} \left[ 1 
+{1\over 3}\left( {1\over 2}g_{tt}\, \nabla^j \nabla_j\; g_{tt}^{-1}
+{1\over 6} R - m^2 \right)\,\tau_0^2 \right. \nonumber \\
&&\qquad +{1\over 3}\left. \left( {1\over 6} R_{,k} {{g_{tt}}^{,k}
\over g_{tt}} -{1\over 12} \nabla^j \nabla_j \; R + 4 v_1 - 4 {v_{000}
\over |g_{tt}|} \right) \tau_0^4 \ln (2\tau_0^2) +O(\tau_0^4) + \cdots\right].
\label{eq:QI_expansion}
\end{eqnarray}
In the limit of $\tau_0 \rightarrow 0$, the dominant term of the above
expression reduces to 
\begin{equation}
\Delta\hat\rho \geq - {3\over 32 \pi^2 \tau_0^4} \, ,\label{eq:asympt_flat}
\end{equation}
which is the quantum inequality in four-dimensional Minkowski space
\cite{F&Ro95,F&Ro97}.  Thus, the term in the square brackets in 
Eq.~(\ref{eq:QI_expansion}) is the short sampling time expansion of
the ``scale'' function \cite{Pfen97a}, and does indeed reduce to 1 in the
limit of the sampling time tending to zero.  We can ask in what range
can we consider a curved spacetime to be ``roughly'' flat?  The condition
is that the correction terms should be small compared to one, i.e.
\begin{equation}
\tau_0 \ll \left| {1\over 2}g_{tt}\, \nabla^i \nabla_i\; g_{tt}^{-1}
+{1\over 6} R - m^2 \right|^{-1/2}\, .\label{eq:little_t}
\end{equation}
Each of the three terms on the right-hand-side of this relation
have different significance.
The $m^2$ term simply reflects the fact that for a massive scalar
field, Eq.~(\ref{eq:QI_expansion}) is valid only when the sampling time
is small compared to the Compton time.  If we are interested in
the massless scalar field, this term is absent.  The scalar curvature
term, if it is dominant, indicates that the flat space inequality is
valid on scales small compared to the local radius of curvature.
This was argued on the basis of the equivalence principle in
Refs.~\cite{F&Ro96,Pfen97b,E&Ro97}, but is now given a more rigorous
demonstration.  The most mysterious term in Eq.~(\ref{eq:little_t})
is that involving $g_{tt}$.  Typically, this term dominates when the
spacetime contains a horizon, and the observer is at rest near
the horizon.  In this case, the horizon would count as a boundary,
so Eq.~(\ref{eq:little_t}) is requiring that $\tau_0$ be small
compared to the proper distance to the boundary

In the particular case of $|g_{tt}| = 1$, we have $\tau_0 = t_0$ and 
Eq.~(\ref{eq:QI_expansion}) reduces to
\begin{equation}
\Delta\hat\rho \geq -{3\over 32 \pi^2 t_0^4} \left[ 1 
+{1\over 3}\left({1\over 6} R - m^2 \right)\,t_0^2 +{1\over 3}
\left( -{1\over 12} \nabla^i \nabla_i \; R + 4  v_1 - 4 v_{000}
\right) t_0^4 \ln (2 t_0^2) + \cdots\right].
\label{eq:gtt_equals_1}
\end{equation}
This result has also been obtained by Song \cite{Song97}, who uses
a heat kernel expansion of the Green's function to develop a short
sampling time expansion.  We can now apply this for a massless scalar
field in the four-dimensional static Einstein universe. The metric
is given by
\begin{equation}
ds^2 = - dt^2 + a^2 \left[ d\chi^2 + \sin^2 \chi ( d\theta^2 +
\sin^2 \theta d\varphi^2 ) \right],
\end{equation}
and the scalar curvature $R = 6/a^2$ is a constant. It can be shown
that $v_1 - v_{000} = 1/8a^4$.  This leads to a quantum inequality
in Einstein's universe of the form
\begin{equation}
\Delta\hat\rho \geq -{3\over 32 \pi^2 t_0^4} \left[
1+ {1\over 3} \left({t_0\over a}\right)^2 + {1\over 3}
\left({t_0\over a}\right)^4 \ln ( t_0 / a) + O({t_0^4 \over a^4})
+ \cdots\right].
\label{eq:QI_Einstein_short}
\end{equation}
In Ref.~\cite{Pfen97a}, an  exact quantum inequality valid for all
$t_0/a$ was derived.  In the limit $t_0 \ll a$, this inequality
agrees with Eq.~(\ref{eq:QI_Einstein_short}).  Similarly, the exact
inequality for the static, open Robertson-Walker universe was obtained
in Ref.~\cite{Pfen97a}, and in the limit $t_0 \ll a$ agrees with
Eq.~(\ref{eq:gtt_equals_1}).

\section{Quantum Inequalities near Planar Mirrors}\label{sec:planar}

\subsection{Single Mirror}
Consider four-dimensional Minkowski spacetime which has a perfectly
reflecting boundary at $z=0$, located in the $x-y$ plane, at which
we require the scalar field to vanish.  The two-point function can be
found by using the standard Feynman Green's function in Minkowski space, 
\begin{equation}
G^{(0)}_F(x,x') ={-i\over 4\pi^2 [ (x-x')^2 + (y-y')^2 +(z-z')^2 -(t-t')^2]},
\end{equation}
and applying the method of images to find the required Green's function
when the boundary is present. For a single conducting plate one finds
\begin{eqnarray}
G_F(x,x') &=& {-i\over 4\pi^2}\left[ {1\over{ (x-x')^2 + (y-y')^2 +(z-z')^2
-(t-t')^2}} \right.\nonumber\\
&&\qquad - \left. {1\over{ (x-x')^2 + (y-y')^2 +(z+z')^2 -(t-t')^2}}\right]
\, .
\end{eqnarray}
If we Euclideanize by allowing $t\rightarrow -it_0$, $t'\rightarrow it_0$
and then take $x' \rightarrow x$, we find
\begin{equation}
G_E(2t_0) = {1\over 16\pi^2}\left( {1\over t_0^2} -
{1\over {t_0^2 + z^2}} \right) .
\end{equation}
In addition, the Euclidean box operator is given by
\begin{equation}
\Box_E = \partial^2_{t_0} +\partial^2_x+\partial^2_y+\partial^2_z.
\end{equation}
It is easily shown that the quantum inequality is given by
\begin{equation}
\Delta\hat\rho \geq -{1\over 4}\, \Box_E \, G_E(2t_0)
= -{3\over 32\pi^2\, t_0^4} + {1\over 16\pi^2 (t_0^2 + z^2)^2}\, .
\end{equation}
The first term of this inequality is identical to that for Minkowski
space.  The second term represents the effect of the mirror on the
quantum inequality. For the minimally coupled scalar field there is
a non-zero, negative vacuum energy density which diverges
as one approaches the mirror. Adding this vacuum term to both the left
and right-hand sides of the above expression allows us to find the
renormalized quantum inequality for this spacetime,
\begin{equation}
\hat\rho_{Ren.} \geq -{3\over 32\pi^2\, t_0^4} +
{1\over 16\pi^2 (t_0^2 + z^2)^2} - {1\over 16\pi^2  z^4}\, .
\end{equation}
There are two limits in which the behavior of the renormalized
quantum inequality can be studied.  First consider $z \gg t_0$.
In this limit, the correction term due to the mirror, and the
vacuum energy very nearly cancel and one finds that the quantum
inequality reduces to
\begin{equation}
\hat\rho_{Ren.} \geq -{3\over 32\pi^2\, t_0^4}.
\end{equation}
This is exactly the expression for the quantum inequality in Minkowski
spacetime.  Thus, if an observer samples the energy density on time
scales which are small compared to the light travel time to the boundary,
then the Minkowski space quantum inequality is a good approximation.

The other important limit is when $z \ll t_0$.  This is the case
for observations made very close to the mirror, but for very long
times.  The quantum inequality then reduces to
\begin{equation}
\hat\rho_{Ren.} \geq -{1\over 16\pi^2\, z^4}.
\end{equation}
Here, we see that the quantum field is satisfying the quantum
averaged weak energy condition.  Recall that throughout the present
paper, we are concerned with observers at rest with respect to the
plate.  If the observer is moving and passes through the plate,
then it is necessary to reformulate the quantum inequalities in
terms of sampling functions with compact support \cite{F&P&R97}.
It should be noted that the divergence of the vacuum energy on the
plate is due to the unphysical nature of perfectly reflecting
boundary conditions.  If the mirror becomes transparent at high
frequencies, the divergence is removed.  Even if the mirror is 
perfectly reflecting, but has a nonzero position uncertainty, the
divergence is also removed \cite{F&SV97b}.

\subsection{Two Parallel Plates}

Now let us consider the case of two parallel plates, one located in
the $z = 0$ plane and another located in the $z=L$ plane.  We are 
interested in finding the quantum inequality in the region between
the two plates, namely $0\leq z \leq L$.  We can again use the method
of images to find the Green's function.  In this case, not only do we
have to consider the reflection of the source in each mirror, but 
we must also take into account the reflection of one image in the
other mirror, and then the reflection of the reflections. This
leads to an infinite number of terms that must be summed to find the
exact form of the Green's function.  If we place a source at
$(t', x', y', z')$, then there is an image of the source  at
$(t', x', y', -z')$ from the mirror at $z=0$ and a second
image at $(t', x', y', 2L-z')$ from the mirror at $z=L$.  Then, we
must add the images of these images to the Green's function, continuing
ad infinitum for every pair of resulting images.  If we use the notation
\begin{equation}
G_F(z, a \pm z') \equiv G^{(0)}_F(t, x , y, z; t', x', y', a \pm z')
\end{equation}
then we can write the Green's functions between the plates as
\begin{eqnarray}
G(x,x') &=& G_F(z, z') - G_F(z, -z') + \sum_{n=1}^\infty\left[
G_F(z, 2nL+z')\right. \nonumber\\
&&\left. - G_F(z, -2nL-z') + G_F(z, -2nL+z') - G_F(z, 2nL-z') \right]
\end{eqnarray}
Again, we Euclideanize as above, and let the spatial separation between
the source and observer points go to zero, we find
\begin{eqnarray}
G_E(2t_0) &=& {1\over 16\pi^2}\left( {1\over t_0^2} -{1\over {t_0^2 + z^2}}
\right) + {1\over 16\pi^2} \sum_{n=1}^\infty\left[{2\over {t_0^2 + (nL)^2}}
\right.\nonumber \\
&& \quad -\left.{1\over {t_0^2 + (nL+z)^2}}-{1\over {t_0^2 +(nl-z)^2}}\right]
\end{eqnarray}
It is now straightforward to find the quantum inequality,
\begin{eqnarray}
\Delta\hat\rho &\geq& -{3\over 32\pi^2\, t_0^4}\, + 
{1\over 16\pi^2 (t_0^2 + z^2)^2} + {1\over 16\pi^2 }\sum_{n=1}^\infty\left\{
{(nL)^2 - 3t_0^2 \over [t_0^2 + (nL)^2]^3} \right.\nonumber\\
&&\quad\left. + {1\over [t_0^2 + (nL+z)^2]^2}
+ {1\over [t_0^2 + (nL-z)^2]^2} \right\}.
\end{eqnarray}
We again have that the first
term in the above expression is identical to that found for Minkowski
space.  The second term is the modification of the quantum inequality
due to the mirror at $ z = 0$.  The modification due to the presence
of the second mirror is contained in the summation, as well as all
of the multiple reflection contributions. When the Casimir vacuum
energy, given by \cite{Fulling}
\begin{equation}
\rho_{vac.} = -{\pi^2\over 48 L^4}\,{3-2\sin^2(\pi z/L) \over 
\sin^4(\pi z/L)}\, - \, {\pi^2 \over 1440 L^4}\, ,
\end{equation}
is added back into this equation for renormalization, we
find, as we did with a single mirror, that close to either of the
mirror surfaces the vacuum energy comes to dominate and the quantum
inequality becomes extremely weak. 

\section{Spacetimes with Horizons}\label{sec:horizons}

We will now change from flat spacetimes with boundaries
to spacetimes in which there exist horizons.  We
will begin with the two-dimensional Rindler spacetime
to develop the quantum inequality for uniformly accelerating
observers.  For these observers, there
exists a particle horizon along the null rays $x = \pm t$ 
(see Fig.~\ref{fig:rindler}).

We will then look at the static coordinatization of 
de~Sitter spacetime.  Again there exists a particle horizon 
in this spacetime, somewhat similar to that of the Rindler
spacetime. The two problems differ somewhat by the fact that
Rindler space is flat while the de~Sitter spacetime
has constant, positive spacetime curvature.
   
\subsection{Two-Dimensional Rindler Spacetime}
\begin{figure}
\begin{center}
\leavevmode\epsffile{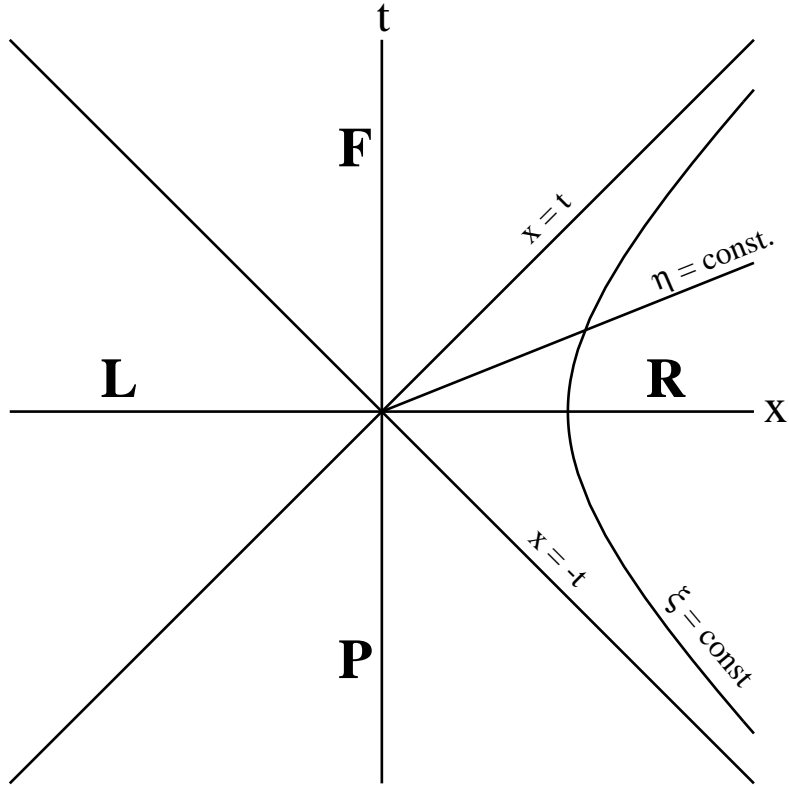}
\end{center}
\caption[Rindler coordinatization of 2-D Minkowski spacetime]
{ A plot of the Rindler coordinatization of two-dimensional Minkowski
spacetime.  The time coordinate $\eta =$ constant are straight lines
passing through the origin, while the space coordinate $\xi =$ constant
are hyperbolae.  The Minkowski spacetime is covered by four separate
coordinate patches, labeled by {\bf L}, {\bf R}, {\bf F} and {\bf P}.
The two null rays ($x=t$ and $x=-t$) act as horizons. }
\label{fig:rindler}
\end{figure}

We begin with the usual two-dimensional Minkowski metric
\begin{equation}
ds^2 =-dt^2+dx^2 .
\end{equation}
Now let us consider an observer who is moving with constant acceleration.
We can transform to the observer's rest frame (Sec.~4.5 of \cite{Brl&Dv})
by 
\begin{eqnarray}
t &=& a^{-1} e^{a\xi} \sinh a\eta\\
x &=& a^{-1} e^{a\xi} \cosh a\eta, 
\end{eqnarray}
where $a$ is a constant related to the acceleration by
\begin{equation}
a\;e^{-a\xi} = {\rm proper\ acceleration}.
\end{equation}
The metric in the rest frame of the observer is then given by
  \begin{equation}
ds^2 = e^{2a\xi}( -d\eta^2 + d\xi^2 )\label{eq:rindler_metric}
\end{equation}
The accelerating observers coordinates $(\eta,\xi)$ only cover
one quadrant of Minkowski spacetime, where $x>|t|$.  This is
shown in Fig.~\ref{fig:rindler}.  Four different coordinate
patches are required to cover all of Minkowski spacetime in the
regions labeled {\bf L}, {\bf R}, {\bf F} and {\bf P}.  For the
remainder of the paper we will be working specifically in the
left and right regions, labeled {\bf L} and {\bf R} respectively.
In these two regions, uniformly
accelerating observers in Minkowski spacetime can be represented
by observers at rest at constant $\xi$ in Rindler coordinates, as
shown by the hyperbola in Fig.~\ref{fig:rindler}.

The massless scalar wave equation in Rindler spacetime is given by
\begin{equation}
e^{-2a\xi} \left(-{d^2\over d\eta^2} + {d^2\over d\xi^2}\right)
\phi(\eta ,\xi) = 0
\end{equation}
which has the positive frequency mode function solutions
\begin{equation}
f_k(\eta ,\xi)= (4\pi\omega)^{-1/2}\;e^{ik\xi \pm i\omega\eta}.
\end{equation}
Here $-\infty<k<\infty$ and $\omega = |k|$. The plus and minus
signs correspond to the left or right Rindler wedges, respectively.
Using the above mode functions, we can expand the general solution
as
\begin{equation}
\phi(\eta ,\xi) = \int_{-\infty}^\infty dk \left[ b_k^L f_k(\eta ,\xi) +
{b_k^L}^\dagger f_k^*(\eta ,\xi) + b_k^R f_k (\eta ,\xi) + 
{b_k^R}^\dagger f_k^*(\eta ,\xi) \right],
\end{equation}
where ${b_k^L}^\dagger$ and ${b_k^L}$ are the creation and annihilation
operators in the left Rindler wedge and similarly ${b_k^R}^\dagger$ and
${b_k^R}$ in the right Rindler wedge.  We also need to define two
vacua, $|0_L\rangle$ and $|0_R\rangle$ with the properties
\begin{equation}
{b_k^L}^\dagger|0_R\rangle ={b_k^R}^\dagger|0_L\rangle =
b_k^L |0_L\rangle = b_k^L |0_R\rangle =b_k^R |0_L\rangle 
=b_k^L |0_R\rangle  = 0.
\end{equation}
The Rindler particle states are then excitations above the vacuum given by
\begin{eqnarray}
|\{1_k\}_L\rangle &=& {b_k^L}^\dagger|0_L\rangle\\
|\{1_k\}_R\rangle &=& {b_k^R}^\dagger|0_R\rangle.
\end{eqnarray} 
With this in hand, we can find the two-point function in either the
left or right hand regions. Let us consider the right hand region,
where
\begin{eqnarray}
G^+(x,x') &=& \langle 0_R| \phi(x) \phi(x') |0_R\rangle\\
&=& \int_{-\infty}^\infty dk \, f_k(x) f_k^*(x')\\
&=& {1\over 4\pi} \int_{-\infty}^\infty {dk\over\omega}
e^{ik (\xi-\xi') - i\omega (\eta-\eta')}
\end{eqnarray}
To find the Euclidean two-point function required for the
quantum inequality, we first allow the spatial separation
to go to zero and then take $(\eta - \eta') \rightarrow -2i\eta_0$,
yielding
\begin{equation}
G_E(2\eta_0) = {1\over 2\pi}\int_0^\infty {d\omega\over\omega}
e^{-2\omega\eta_0}.
\end{equation}
In two-dimensions, the Euclidean Green's function for the massless
scalar field has an infrared divergence as can be seen from the form
above, in which the integral is not well defined in the limit of
$\omega\rightarrow 0$.  However, in the process of finding the 
quantum inequality we act on the Green's function with the Euclidean
box operator. If we first take the derivatives of the Green's function,
and then carry out the integration, the result is well defined for all
values of $\omega$. In Rindler space, the Euclidean box operator is 
given by
\begin{equation}
\Box_E = e^{-2a\xi} \left({d^2\over d\eta^2} + {d^2\over d\xi^2}\right).
\end{equation}
It is now easy to solve for the quantum inequality, 
\begin{equation}
\Delta\hat\rho \geq -{1\over4} \Box_E G_E(2\eta_0)
= -{1\over 2\pi} e^{-2a\xi} \int_o^\infty d\omega\;\omega
e^{-2\omega\eta_0}
= -{1\over 8\pi \left(e^{a\xi}\,\eta_0\right)^2}.
\end{equation}
However, the coordinate time $\eta_0$ is related to the observer's proper
time by
\begin{equation}
\tau_0 = e^{a\xi}\eta_0,
\end{equation}
allowing us to rewrite the quantum inequality in a more covariant
form,
\begin{equation}
\Delta\hat\rho\geq -{1\over 8\pi\tau_0^2}\,.\label{eq:2D_QI}
\end{equation}
This is exactly the same form of the quantum inequality as found
in two-dimensional Minkowski spacetime \cite{F&Ro95,F&Ro97}. We will
see in Sec.~\ref{sec:2D_Black_holes} that this is a typical property
of static two-dimensional spacetimes.  This arises because in two
dimensions all static spacetime are conformal to one another.
However, the renormalized quantum inequalities are not identical 
in different spacetimes because of differences in the vacuum energies.

\subsection{de~Sitter Spacetime}
Let us now consider four-dimensional de~Sitter spacetime.  The scalar
field quantum inequality, Eq.~(\ref{eq:QI}), assumes a timelike Killing
vector, so it will be convenient to use the static parameterization
of de~Sitter space,
\begin{equation}
ds^2 = -(1-{r^2\over \alpha^2})\,dt^2 + (1-{r^2\over \alpha^2})^{-1}
dr^2 + r^2(d\theta^2 + \sin^2\theta\;d\varphi^2)\; .
\end{equation}
There is a particle horizon at $r = \alpha$ for an observer sitting
at rest at $ r = 0$.  The coordinates take the values, $0 \leq r < 
\alpha$, $0\leq\theta <\pi$ and $0\leq\varphi <2\pi$.  It should
be noted that this choice of metric covers one quarter of de~Sitter 
spacetime.     

The scalar wave equation is
\begin{equation}
\left(1-{r^2\over \alpha^2}\right)^{-1} \partial_t^2 \phi - {1\over r^2}
\partial_r\left[ r^2 \left(1-{r^2\over \alpha^2}\right) \partial_r 
\right]\phi -{1\over r^2}\left[ {1\over \sin\theta}\partial_\theta \left(
\sin\theta\,\partial_\theta \right) + {1\over \sin^2\theta}
\partial_\varphi^2 \right]\phi + \mu^2\phi  = 0\,.
\end{equation}
The unit norm positive frequency mode functions are found 
\cite{Lohiya,Lapedes,Higuchi,Sato94,Kaiser} to be of the form
\begin{equation}
\hat\phi_{\omega,l,m}(t,r,\theta,\varphi) = {1\over\sqrt{4\pi\alpha^3
\omega}}\;f_\omega^l(z) \;{\rm Y}_{lm}(\theta,\varphi)
\;e^{-i\omega t}
\end{equation}
where $z = r/\alpha$ is a dimensionless length, the ${\rm Y}_{lm}
(\theta,\varphi)$'s are the standard spherical
harmonics and the mode labels $l$ and $m$ take the values  $l = 0, 1,
2, \cdots$ and $-l\leq m\leq l$.  The radial portion of the the
solution is given by   
\begin{equation}
f_\omega^l(z) = {\Gamma(b^+_l)\Gamma(b^-_l) \over \Gamma(l+{3\over 2})
\Gamma(i\alpha\omega)} \; z^l \; (1-{z^2})^{i\alpha\omega/ 2}\; 
{\rm F}(\,b^-_l, \,b^+_l;\, l+{3\over 2} ;\, z^2),
\end{equation}
where ${\rm F}(\alpha,\beta ;\gamma ; z)$ is the hypergeometric 
function \cite{Gradshteyn} and
\begin{equation}
b^\pm_l = {1/2}\left(l + {3/2} + i\alpha\omega \pm\sqrt{{9/4}
-\alpha^2\mu^2}\right).\label{eq:b_pm}
\end{equation}
We can then express the two-point function as
\begin{equation}
G(x,x') = \sum_{lm}\int_0^\infty\,dk {1\over 4\pi\alpha^2 k}
{f_k^l}^*(z) \; f_k^l(z') \;{\rm Y}_{lm}^*(\theta,\varphi)\; 
{\rm Y}_{lm}(\theta',\varphi') \; e^{ik(t-t')/a},
\end{equation} 
where $k \equiv \alpha\omega$.  Now if we Euclideanize according to
Eq.~(\ref{eq:Euclideanize}) and set the spatial separation of the
points to zero, we may make use of the  addition theorem for the
spherical harmonics \cite{Jackson}, 
\begin{equation}
\sum_{m=-l}^l \left| {\rm Y}_{lm}(\theta,\varphi) \right|^2 
= {{2l+1}\over 4\pi}\; ,\label{eq:sum_rule}
\end{equation}
to find the Euclidean Green's function
\begin{equation}
G_E = {1\over 16\pi^2\alpha^2}
\sum_l \int_0^\infty dk {(2l+1)\over k} \left|{\Gamma(b^+_l)\Gamma(b^-_l)
\over \Gamma(l+{3\over 2}) \Gamma(ik)} \right|^2
 z^{2l} \, \left|{\rm F}(b^-_l, b^+_l ; l+{3\over 2} ;\, z^2)
\right|^2  e^{-2kt_0 / \alpha}\, . \label{eq:deSitterGreens}
\end{equation}
This is independent of the angular coordinates, as expected,
because de~Sitter space is isotropic. We now need the Euclidean box
operator.  Because of the angular independence of the Green's function,
it is only necessary to know the temporal and radial portions of the
box operator.  One finds that the energy density inequality,
Eq.~(\ref{eq:QI}), becomes
\begin{equation}
\Delta\hat \rho \geq  -{1\over 4}\left\{ {1\over(1-z^2)}
\partial_{t_0}^2 + {1\over\alpha^2 z^2} \partial_z
\left[ z^2 \,(1-z^2) \partial_z \right] \right\}\;
G_E({\bf x},-t_0;{\bf x},+t_0).
\label{eq:open_form}
\end{equation}

The temporal derivative term in Eq.~(\ref{eq:open_form}) will simply
bring down two powers of $k/\alpha$. Using the properties of the
 hypergeometric function, it can be shown that
\begin{equation}
\left|{\rm F}(b^-_l, b^+_l; l+{3\over 2} ;\, z^2) \right|^2 =
(1-z^2)^{ik} \;{\rm F}^2(b^-_l, b^+_l ; l+{3\over 2} ;\, z^2),
\end{equation}
from which we can take the appropriate spatial derivatives. 
If we allow $z \rightarrow 0$, then we have $F \rightarrow 1$
and only the $l=0$ terms will contribute in the time derivative part of
Eq.~(\ref{eq:open_form}).  For the radial derivative, one may show    
\begin{equation}
\lim_{z\rightarrow 0}{1\over z^2} \partial_z \left\{ z^2 \,(1-z^2) \partial_z 
\left[ z^{2l}(1-z^2)^{ik} {\rm F}^2( z^2)
\right]\right\}
=
\left\{ \matrix{  
         2(\alpha^2\mu^2-k^2)&{\rm for}\; l=0\cr
         6&{\rm for}\;l=1\cr
         0&{\rm otherwise.}} \right.
\end{equation}
Using these results, we find for the observer at $r = 0$, that
\begin{equation}
\Delta\hat\rho \geq -{1\over 8\pi^4 \alpha^4 } \int_0^\infty dk 
\sinh(\pi k) \left[
(k^2+\alpha^2\mu^2)\left|\Gamma(b^-_0)\,\Gamma(b^+_0)\right|^2+
4\left|\Gamma(b^-_1)\,\Gamma(b^+_1)\right|^2\right] e^{-2 t_0 k / \alpha}.
\end{equation}
There are two cases for which the right hand side can be evaluated
analytically, $\mu = 0$ and $\mu = \sqrt{2}/\alpha$. For $\mu = 0$,
we have
\begin{eqnarray}
\Delta\hat\rho &\geq& -{1\over 8\pi^4 \alpha^4 } \int_0^\infty dk 
\sinh(\pi k) \left[
k^2\left|\Gamma(i{k\over 2})\,\Gamma({3\over 2}+i{k\over 2})\right|^2
\right.\nonumber\\ & &\qquad\qquad\left.
+4\left|\Gamma({1\over 2}+i{k\over 2})\,\Gamma(2+i{k\over 2})
\right|^2\right] e^{-2 t_0 k / \alpha}\\
&=&-{1\over 8\pi^2 \alpha^4 }\int_0^\infty dk (2k^3+5k)
e^{-2 t_0 k / \alpha}\\
&=&-{3\over 32\pi^2 t_0^4 }\left[ 1 + {5\over3}
\left({t_0\over\alpha}\right)^2 \right]\, ,
\label{eq:minimalDeSitter}\end{eqnarray}
where we have made use of the identities 
\begin{eqnarray}
\left | \Gamma(ik/2) \right |^2 &=&{\pi\over k/2 \sinh(\pi k /2)}, \\
\left | \Gamma(1/2 + ik/2) \right |^2 &=& {\pi\over \cosh(\pi k /2)},\\
\left | \Gamma(1 + ik/2) \right |^2 &=& {\pi k/2 \over \sinh(\pi k/2)},\\
\left | \Gamma(3/2 + ik/2) \right |^2 &=& {\pi\over 2} \left(1+k^2\right)
{\cosh(\pi k /2) \over \cosh(\pi k) +1},
\end{eqnarray}
and
\begin{equation}
\left | \Gamma(2 + ik/2) \right |^2 = {\pi\over 4} k\,\left(4+k^2\right)
{\sinh(\pi k /2) \over \cosh(\pi k) -1}\, .
\end{equation}
Similarly for $\mu = \sqrt{2}/\alpha$, we find
\begin{equation}
\Delta\hat\rho \geq -{3\over 32\pi^2 t_0^4 }\left[ 1 + 
\left({t_0\over\alpha}\right)^2 \right].
\label{eq:ConformalDeSitter}\end{equation}
We can compare these results with the short sampling time approximation
from Sec.~\ref{sec:expans}.  Solving for the necessary geometrical
coefficients, we find
\begin{eqnarray}
v_{000} &=& \left( {29\over 60}{1\over\alpha^4} - {1\over 4}{\mu^2\over
\alpha^2}\right) |g_{tt}|\, ,\\[8pt]
v_1 & = & {29\over 60}{1\over\alpha^4} - {1\over 2}{\mu^2\over
\alpha^2} + {1\over 8} \mu^4 \, , \\[8pt]
{1\over 2}g_{tt}\nabla^j\nabla_j\, g_{tt}^{-1} &=& {1\over\alpha^2} 
{\left(3-r^2/\alpha^2\right) \over \left(1-r^2/\alpha^2\right)}
\, . \end{eqnarray}
The general short time expansion, Eq.~(\ref{eq:QI_expansion}), now becomes
\begin{eqnarray}
\Delta\hat\rho &\geq & - {3\over 32 \pi^2 \tau_0^4} \left\{ 1+ {1\over 3}
\left[ {2\over\alpha^2} - \mu^2 + {1\over\alpha^2}{(3-r^2/\alpha^2) 
\over (1-r^2/\alpha^2)} \right] \, \tau_0^2 \right. \nonumber \\
&& \qquad\qquad\left. + {\mu^2\over 6}\left(
\mu^2 - {2\over\alpha^2}\right)\, \tau_0^4 \, \ln (2 \tau_0^2 / 
\alpha^2) + O(\tau_0^4) + \cdots\right\}, \end{eqnarray}
where $\tau_0 = (1 -r^2/\alpha^2 )^{1/2} t_0$.
If $r = 0 $ and $\mu$ takes the values $0$ or $\sqrt{2}/\alpha$, this
agrees with Eqs.~(\ref{eq:minimalDeSitter}) or (\ref{eq:ConformalDeSitter}),
respectively.  Note that this small $t_0$ expansion is valid for all radii, 
$0 \leq r < \alpha$.  We can also find the proper sampling time from
Eq.~(\ref{eq:little_t}) for which this expansion is valid,
\begin{equation} 
\tau_0 \ll \tau_m \equiv \alpha \sqrt{{1-r^2/\alpha^2} \over
{5 - 3r^2/\alpha^2}}.
\end{equation}    
For the observer sitting at the origin of the coordinate system,
$\tau_0 \ll \alpha/ \sqrt{5}$.  This is the scale on which
the spacetime can be considered ``locally flat''. For observers at
$r > 0$, who do not move on geodesics, $\tau_m$ decreases and
approaches zero as $r \rightarrow \alpha$:  
\begin{equation}
\tau_m \sim \sqrt{\alpha (\alpha-r)}\qquad , r\rightarrow\alpha.
\end{equation}
Note that the proper distance to the horizon from radius r is
\begin{eqnarray}
\ell &=& \int_r^\alpha {dr' \over \sqrt{1 -r'^2 /\alpha^2 }} =
\alpha\left[ \pi/2 - \arcsin (r/\alpha)\right]\\
&\sim&\sqrt{2\alpha (\alpha-r)} ,\qquad \mbox{ as } r\rightarrow\alpha.
\end{eqnarray}
Thus, for observers close to the horizon, if the sampling time is
small compared to this distance to the horizon, $\tau_0 \ll \ell$,
then $\tau_0 \ll \tau_m$, and the short time expansion is valid. 

We can also obtain a renormalized quantum inequality
for the energy density at the origin for the case  $\mu = \sqrt{2}
\alpha$.  By the addition of the vacuum energy to both  sides of
Eq.~(\ref{eq:ConformalDeSitter}) one finds
\begin{equation}
\hat\rho_{Ren.} \geq -{3\over 32\pi^2 t_0^4 }\left[ 1 + 
\left({t_0\over\alpha}\right)^2 \right] - {1\over 960\pi^2\alpha^4}.
\end{equation}
We can now predict what will happen in the infinite
sampling time limit of the renormalized quantum inequality for any
observer's position. We know from Eqs.~(\ref{eq:deSitterGreens})
and (\ref{eq:open_form}) that the difference inequality will always
go to zero, yielding a QAWEC in static de~Sitter space of
\begin{equation}
\lim_{t_0\rightarrow\infty} {t_0\over\pi} \int_{-\infty}^\infty 
{\langle T_{00}u^0 u^0 \rangle_{Ren.} \over  t^2 + t_0^2} dt \geq
{1\over 480\pi^2\alpha^4}\left[ -{\alpha^2\over (\alpha^2-r^2)}+
{1\over2}(1-{r^2\over\alpha^2}) \right].
\label{eq:QAWECdeSitter}
\end{equation}
We immediately see that for a static observer who is arbitrarily close
to the horizon in de~Sitter spacetime, the right hand side of
Eq.~(\ref{eq:QAWECdeSitter}) becomes extremely negative, and diverges
on the horizon itself.  This is similar to the behavior found
for static observers located near the perfectly reflecting mirror
discussed earlier.


\section{Black Holes}\label{sec:black_holes}
We now turn our attention to an especially
interesting spacetime in which quantum inequalities can be
developed, the exterior region of a black hole in two
and four dimensions.     
 
\subsection{Two-Dimensional Black Holes}\label{sec:2D_Black_holes}
Let us consider the metric
\begin{equation}
ds^2 = - C(r)\,dt^2 + C(r)^{-1}\, dr^2\, ,
\end{equation}
where $C(r)$ is a function chosen such that $C\rightarrow1$ and 
$\partial C / \partial r \rightarrow 0$ as $r\rightarrow \infty$.
Additionally, there is an event horizon at some value $r_0$
where $C(r_0) = 0$.  For example, in the Schwarzschild
spacetime, $C(r) = 1 -2Mr^{-1}$, there is a horizon at $r = 2M$.
Another choice for $C$ is that of the Reissner-Nordstrom black
hole, where $C(r) = 1 -2Mr^{-1} + Q^2 r^{-2}$.  In general,
we will leave the function $C$ unspecified for the remainder
of the derivation.  The above metric leads to the massless,
minimally coupled scalar wave equation
\begin{equation}
-{1\over C(r)}\partial_t^2\phi(r,t) + \partial_r\left[C(r)
\partial_r \phi(r,t) \right] = 0\, .
\end{equation}
Unlike in 4-dimensions, the 2-dimensional wave equation can be
analytically solved everywhere.  If we use the standard definition
of the $r^*$ coordinate,
\begin{equation}
r^* \equiv \int {dr\over C(r)}\, ,
\end{equation}
then it is convenient for us to take as the definition
of the positive frequency mode functions
\begin{equation}
f_k(r,t) = i\left( 4\pi\omega \right)^{-1/2} \;
e^{ikr^*-i\omega t}\qquad \omega = |k|,
\end{equation}
where $-\infty < k < \infty$.  

The problem of finding the quantum inequality simply reduces to using the
mode functions to find the Euclidean Green's function.  We have 
\begin{equation}
G_E(2t_0) = \int_{-\infty}^\infty dk\;\left|{i\over \sqrt{4\pi\omega}}
e^{ikr^*}\right|^2 \, e^{-2\omega t_0}
= {1\over 2\pi} \int_0^\infty d\omega\;\omega^{-1}
e^{-2\omega t_0} 
\end{equation}
As in the case of two-dimensional Rindler space, the Euclidean
Green's function has an infrared divergence.  We can again apply
the Euclidean box operator first and then do the integration to
obtain the quantum inequality, 
\begin{equation}
\Delta\hat\rho \geq -{1\over 2\pi C(r)}\int_0^\infty d\omega\;
\omega\,e^{-2\omega t_0}  =  -{1\over 8\pi C(r) t_0^2}\, . 
\end{equation}
However, the observer's proper time is related to the coordinate time
by $\tau = C(r)^{1/2} t$, such that we can write the difference
inequality as
\begin{equation}
\Delta\hat\rho \geq  -{1\over 8\pi \tau_0^2}.
\end{equation}
This is the same form as found for two-dimensional Minkowski and
Rindler spacetime. This is the expected result because all
two-dimensional static spacetimes are conformal to one and other.
For an extensive treatment of quantum inequalities in two-dimensional
Minkowski spacetime, see \cite{Flan97}.

This now brings us to the matter of renormalization.  There exist
three candidates for the vacuum state of a black hole: the Boulware
vacuum, the Hartle-Hawking vacuum, and the Unruh vacuum.  However
the derivation of the difference inequality relies on the mode
functions being defined to have positive frequency with respect to
the timelike Killing vector $\partial_t$, and that the vacuum state
was destroyed by the annihilation operator, i.e.
\begin{equation}
a_k \; |0_k\rangle = 0 \qquad{\rm for\; all}\; k\, .
\end{equation}
In Schwarzschild spacetime, this defines the Boulware vacuum.
Thus, we can solve for the renormalized quantum inequality,
\begin{equation}
\hat\rho_{Ren.}\equiv{t_0\over\pi} \int_{-\infty}^\infty 
{\langle{T_{tt}} \rangle_{Ren.}\over{t^2+t_0^2}}dt
\geq  -{1\over 8\pi \tau_0^2} + \rho_{B}(r)\; .
\end{equation}
The Boulware vacuum energy density in two-dimensions
for the Reissner-Nordstrom black hole, is given explicitly by 
(see Sec.~8.2 of\cite{Brl&Dv})
\begin{equation}
\rho_{B}(r) = {1\over 24\pi}\left(1 -{2M\over r} + {Q^2\over r^2}
\right)^{-1}\left[ -{4M\over r^3}+
{7 M^2\over r^4} + {6 Q^2\over r^4} - {14 MQ^2\over r^5} +
{5 Q^4\over r^6}\right]
\end{equation}
In the limit $\tau_0 \rightarrow \infty$, one recovers a QAWEC
condition on the energy density
\begin{equation}
\lim_{t_0\rightarrow\infty} {t_0\over \pi} \int_{-\infty}^\infty 
{\langle{T_{tt}/g_{tt}}\rangle_{Ren.}\over t^2+t_0^2} dt
\geq \rho_{B}(r).\label{eq:AWEC_boulware}
\end{equation}
This has the interpretation that the integrated energy density in
an arbitrary  particle state can never be more negative than that
of the Boulware vacuum state. In particular, this will be true for
the Hartle-Hawking and Unruh vacuum states.  

\subsection{Four-Dimensional Schwarzschild Spacetime}

Now let us turn to the four-dimensional Schwarzschild  spacetime with
the metric
\begin{equation}
ds^2 = -\left(1-{2M\over r}\right) dt^2 + \left(1-{2M\over r}\right)^{-1} dr^2
+ r^2\,(d\theta^2 + \sin^2\theta\, d\varphi^2)\, .
\end{equation}
The normalized mode functions for a massless scalar field in the
exterior region ($r>2M$) of Schwarzschild spacetime can be written as 
\cite{DeWitt}
\begin{eqnarray}
\stackrel{\rightarrow}{f}_{\omega l m}(x) &=& (4\pi\omega)^{1/2} 
e^{-i\omega t} \stackrel{\rightarrow}{R}_l(\omega | r)
{\rm Y}_{lm}(\theta,\varphi),\nonumber\\
\stackrel{\leftarrow}{f}_{\omega l m}(x) &=& (4\pi\omega)^{1/2}
e^{-i\omega t} \stackrel{\leftarrow}{R}_l(\omega | r)
{\rm Y}_{lm}(\theta,\varphi),
\end{eqnarray}
where $\stackrel{\rightarrow}{R}_l(\omega | r)$ and 
$\stackrel{\leftarrow}{R}_l(\omega | r)$ are the
outgoing and ingoing solutions to the radial portion
of the wave equation, respectively.  Although they cannot be written
down analytically, their asymptotic forms are
\begin{equation}
\stackrel{\rightarrow}{R}_l(\omega | r) \sim \left\{ 
\matrix{ r^{-1} e^{i\omega r^*} + \stackrel{\rightarrow}{\rm A}_l
         (\omega) r^{-1} e^{-i\omega r^*}\, ,& \qquad r\rightarrow 2M\cr
        {\rm B}_l(\omega) r^{-1} e^{i\omega r^*}\, ,
         & \qquad r\rightarrow \infty \cr}\right.
\end{equation}
for the outgoing modes and
\begin{equation}
\stackrel{\leftarrow}{R}_l(\omega | r) \sim \left\{
\matrix{{\rm B}_l(\omega) r^{-1} e^{-i\omega r^*}\, ,
        & \qquad r\rightarrow 2M\cr
        r^{-1} e^{-i\omega r^*} + \stackrel{\leftarrow}{\rm A}_l(\omega)
        r^{-1} e^{i\omega r^*}\, ,& \qquad r\rightarrow \infty \cr}\right.
\end{equation}
for the ingoing modes.  The normalization factors ${\rm B}_l(\omega)$, 
$\stackrel{\rightarrow}{\rm A}_l(\omega)$ and $\stackrel{\leftarrow}{\rm A}_l
(\omega)$ are the transmission and reflection coefficients
for the scalar field with an angular momentum-dependent potential
barrier.

Now let us consider the two-point function in the Boulware vacuum.  It is
given by 
\begin{eqnarray}
G_B(x,x') &=&  \sum_{lm} \int_0^\infty {d\omega \over 4\pi\omega}
e^{-i\omega(t-t')}\,{\rm Y}_{lm}(\theta,\varphi) {\rm Y}_{lm}^*
(\theta',\varphi') \cr
&& \qquad \times \left[\stackrel{\rightarrow}{R}_l(\omega | r)
\stackrel{\rightarrow}{R}_l^*(\omega | r') +\stackrel{\leftarrow}{R}
_l(\omega | r) \stackrel{\leftarrow}{R}_l^*(\omega | r')\right]\,.
\end{eqnarray}
We are interested in the two-point function when the spatial separation
goes to zero, i.e. letting $r'\rightarrow r$, $\theta' \rightarrow \theta$,
and $\varphi' \rightarrow \varphi$.  We can again make use of the 
addition theorem, Eq.~(\ref{eq:sum_rule}), for the spherical harmonics.
Let us also Euclideanize, by taking $(t-t')\rightarrow
-2it_0$.  The Euclidean two-point function then reduces to
\begin{equation}
G_{BE}(2t_0) = {1\over 16\pi^2} \sum_l \int_0^\infty {d\omega\over\omega}\,
e^{-2t_0}\,(2l+1)\, \left[|\stackrel{\rightarrow}{R}_l(\omega | r)|^2
+|\stackrel{\leftarrow}{R}_l(\omega | r)|^2\right]\,.
\end{equation}
In the two asymptotic regimes, close to the event
horizon of the black hole ($r\rightarrow 2M$), or far from the
black hole ($r\rightarrow\infty$), the
radial portion of the wave equation also satisfies a sum rule.
It was found by Candelas \cite{Candelas} that
\begin{equation}
\sum_{l=0}^\infty (2l+1) |\stackrel{\rightarrow}{R}_l(\omega | r)|^2
\sim \left\{ \matrix{
             4\omega^2(1- 2M/ r)^{-1},& \qquad 
             r\rightarrow 2M \cr
             r^{-2}\sum_{l=0}^\infty (2l+1) |{\rm B}_l(\omega)|^2  ,
             & \qquad r\rightarrow \infty 
            }
     \right.
\end{equation}
and
\begin{equation}
\sum_{l=0}^\infty (2l+1) |\stackrel{\leftarrow}{R}_l(\omega | r)|^2
\sim \left\{ \matrix{ 
             (2M)^{-2}\sum_{l=0}^\infty (2l+1) |{\rm B}_l(\omega)|^2,
             & \qquad  r\rightarrow 2M \cr
             4\omega^2  ,&  \qquad r\rightarrow \infty
            }
     \right.
\end{equation}
with the coefficient ${\rm B}_l(\omega)$ given, in the case
$2M\omega \ll 1$, by \cite{Jens92}
\begin{equation}
{\rm B}_l(\omega)\approx {(l!)^3\over (2l+1)!\,(2l)!}\,
(-4iM\omega)^{l+1}\, . \label{eq:Jenson} \end{equation}
If we insert these relations into the Green's functions,  it is
possible to carry out the integration in $\omega$.  One finds
\begin{equation}
G_{BE}(2t_0) \sim {1\over 16\pi^2} \left[ {1\over(1-2M/ r) t_0^2}
+{1\over 4M^2} \sum_{l=0} {(l!)^6\over [(2l)!]^3}\,
\left({2M\over t_0}\right)^{2l+2}\right] , \qquad r\rightarrow 2M
\label{eg:green_near}\end{equation}
in the near field limit and in the far field limit,
\begin{equation}
G_{BE}(2t_0) \sim {1\over 16\pi^2} \left[ {1\over t_0^2}+ {1\over r^2}
\sum_{l=0}{(l!)^6\over [(2l)!]^3}\,\left({2M\over t_0}\right)^{2l+2}
\right] , \qquad r\rightarrow \infty. \label{eq:green_far}
\end{equation}
We immediately see that the Green's function is independent of the
angular coordinates, as one expects because of spherical symmetry.
Note that the maximum value of $l$ for which the expansion in
Eqs.~(\ref{eg:green_near}) and (\ref{eq:green_far}) can be used
depends upon the order of the leading terms which have been dropped
in Eq.~(\ref{eq:Jenson}).  If this correction is $O\left( (M\omega)^{l+2}
\right)$, then only the $l=0$ terms are significant, as $B_0$ would
then contain subdominant pieces which yield a contribution to
$G_{BE}(2t_0)$ larger than the leading contribution from $B_1$.
In what follows, we will explicitly retain only the $l=0$ contribution.
In order to find the quantum inequality around a black hole we must
evaluate
\begin{equation}
\Delta\hat\rho \geq -{1\over 4}\, \Box_E \; G_E(2t_0).
\end{equation}
However, the only parts of the Euclidean box operator
that are relevant are the temporal and radial terms, i.e.
\begin{equation}
\Box_E \Rightarrow (1-{2M/r})^{-1}\partial_{t_0}^2 + 
r^{-2}\partial_r [r^2(1-{2M/r})\partial_r].
\end{equation}
Upon taking the appropriate derivatives, and using the relation of
the proper time of a stationary observer to the coordinate time:
\begin{equation}
\tau_0 = t_0\,\sqrt{1-{2M/r}}\, , 
\end{equation}
we find that the quantum inequality is given by
\begin{eqnarray}
\Delta \hat\rho &\geq& -{3\over 32\pi^2 \tau_0^4 }\left\{  
{1\over 6} \left(2M\over r\right)^2 \left( \tau_0 \over r \right)^2
\left(1-{2M \over r}\right)^{-1} + 1 +\left(1-{2M \over r}\right)+ 
\right.\nonumber \\
&&\qquad\qquad \left. + O\left[ \left(1-
{2M \over r}\right)^2 \right] + \cdots\right\}\qquad\qquad r\rightarrow 2M
\label{eq:near}
\end{eqnarray}
and
\begin{eqnarray}
\Delta \hat\rho &\geq& -{3\over 32\pi^2 \tau_0^4} \left\{
1 - {2M\over r} + \left({2M\over r}\right)^2 \left[ 1 +{1\over 3}
\left({\tau_0\over r}\right)^2\right]  - \left({2M\over r}\right)^3
\left[ 1 +\left({\tau_0\over r}\right)^2\right] + \right.\nonumber \\
&&\qquad\qquad \left. + O\left[ \left(
{2M\over r}\right)^4\right] +\cdots \right\}
\qquad\qquad r\rightarrow\infty
\label{eq:far}  
\end{eqnarray}
An alternative approach to finding the quantum inequality is to use
the short time expansion from Sect.~\ref{sec:expans}, which yields
\begin{equation}
\Delta \hat\rho \geq -{3\over 32\pi^2 \tau_0^4} - {1\over 16\pi^2\tau_0^2}
\left[{M^2\over  r^4 (1- 2M/ r)} + O(\tau_0^2) + \cdots\right].
\label{eq:shortSchwarz}
\end{equation}

Note that this short time expansion coincides with the first two terms
of the $r\rightarrow 2M$ form, Eq.~(\ref{eq:near}).  This is somewhat
unexpected, as Eq.~(\ref{eq:near}) is an expansion for small $r-2M$ with
$\tau_0$ fixed, whereas Eq.~(\ref{eq:shortSchwarz}) is an expansion for
small $\tau_0$ with $r$ fixed.

We immediately see from Eq.~(\ref{eq:far}) 
that we recover the Minkowski space quantum inequality in the $r
\rightarrow \infty$ limit.   If we consider experiments performed
on the surface of the earth, where the radius of the earth
is several orders of magnitude larger than its equivalent Schwarzschild
radius, then the flat space inequality is an exceptionally good 
approximation. From Eq.~(\ref{eq:little_t}), we can also find the
proper sampling time for which the inequality Eq.~(\ref{eq:shortSchwarz})
holds to be
\begin{equation} 
\tau_0 \ll {r^2\over 2M} \sqrt{2\left(1-{2M\over r}\right)}.
\end{equation}

As was the case in two dimensions, if we allow the sampling time
to go to infinity in the exact quantum inequality, we recover the 
QAWEC, Eq.~(\ref{eq:AWEC_boulware}), for the four-dimensional black
hole.  The QAWEC says that the renormalized energy density for an
arbitrary particle state, sampled over the entirety of the rest
observer's worldline can never be more negative than the
Boulware vacuum energy density.


\section{Summary and Conclusions}
We have shown for static spacetimes that the energy density sampled
for a characteristic time $t_0$ along the worldline of a static
observer is bounded below by the quantum inequality, 
\begin{equation}
\Delta\hat\rho \equiv {t_0\over\pi} \int_{-\infty}^\infty 
{\langle :T_{00}:/g_{tt} \rangle \over  t^2 + t_0^2} dt\geq - {1\over 4}
\left({\partial_{t_0}^2\over g_{tt}} + \nabla^j\nabla_j\right)
\sum_\lambda |U_\lambda({\bf x})|^2 {\rm e}^{-2\omega_\lambda t_0}\, , 
\end{equation}
An observer doing the sampling may observe negative energy
densities.  However, as we have seen in the various examples here
and in previous work \cite{Pfen97a,F&Ro97},  the magnitude of the
sampled negative energy density is bounded below, in four
dimensions, by
\begin{equation}
\Delta\hat\rho \geq - {3\over 32 \pi^2 t_0^4} f(t_0)\, .
\end{equation}
Here, $f(t_0)$ is called the scale function and carries the specific
information about how the quantum inequality is modified from the
flat space form when we are in curved spacetimes.  It has
the general property that when the sampling time of the observation
becomes small, the sampling function $f(t_0) \rightarrow 1$, and we
recover the Minkowski space form of the quantum inequality.  

We may also write the quantum inequality in terms of the Euclidean
box operator and the Euclidean Green's function,
\begin{equation}
\Delta\hat\rho \geq -{1\over 4}\Box_E G_E({\bf x}, -t_0; {\bf x}, +t_0)
\end{equation}
and thus avoid carrying out the sum over all the modes if the Green's
function is already known. If the Green's function in a particular
spacetime is not explicitly known, we can still find the quantum
inequality by using an expansion of the Hadamard form of the Green's
function in the limit of small sampling times.  In Section \ref{sec:expans},
it was shown that the quantum inequality in this limit is given by
Eq.~(\ref{eq:QI_expansion}), which gives the curvature-dependent
corrections to the flat space inequality, Eq.~(\ref{eq:asympt_flat}).
This result confirms the arguments made in Ref.~\cite{F&Ro96} and
further utilized in \cite{Pfen97b,E&Ro97} to the effect that the flat
spacetime quantum inequality may be used in curved spacetimes if the
sampling time is sufficiently short.

In the limit of long sampling time, $t_0\rightarrow \infty$,
one can derive a quantum averaged weak energy
condition, Eq.~(\ref{eq:QAWEC}), which says that the expectation
value of the renormalized energy density for a static observer
sampled for all time is bounded below by the vacuum self energy
of the spacetime.

An exact quantum inequality was found in several examples, including
perfectly reflecting mirrors in flat spacetime, Rindler and de~Sitter
spacetimes and two-dimensional black hole spacetimes.  In all cases,
the short sampling time limit agrees with the general short sampling
time expansion derived in Sec.\ref{sec:expans}.  Approximate forms of
the quantum inequality in four-dimensional Schwarzschild spacetime
were found in the vicinity of the horizon and at large distances.
This inequality places a limit on how much more negative the local
energy density in an arbitrary state may be than that in the 
Boulware vacuum state.

\vspace{18pt} 
\centerline {\bf Acknowledgments}
We would like to thank Thomas A. Roman and Allen Everett for
useful discussions. This research was supported in part by NSF Grant
No.~Phy-9507351 and the John~F.~Burlingame Physics Fellowship Fund.

\newpage


\newpage

\newpage

\end{document}